\documentclass{aa}
\usepackage[varg]{txfonts}

\usepackage{graphicx}
\usepackage{caption}
\usepackage{subcaption}

\usepackage{natbib}
\bibpunct{(}{)}{;}{a}{}{,}
\usepackage[citecolor=blue, linkcolor=blue, urlcolor = black, colorlinks = true]{hyperref}

\usepackage{longtable}
\usepackage{lscape}

\usepackage{verbatim}
\usepackage{makecell}
\usepackage{multirow}
\usepackage{booktabs}

\usepackage{xspace}
\usepackage{comment}
\usepackage{textcomp}
\usepackage[separate-uncertainty=true]{siunitx}
\usepackage[dvipsnames]{xcolor}
\usepackage[normalem]{ulem}

\newcommand{\Msun}{\,M$_{\odot}$\xspace}
\newcommand{\mesa}{\texttt{MESA}\xspace}
\newcommand{\gyre}{\texttt{GYRE}\xspace}
\newcommand{\grad}{$g_{{\rm rad}, i}$\xspace}
\newcommand{\kappar}{$\kappa_{\rm R}$\xspace}

\begin{document}

\title{The impact of radiative levitation on mode excitation of main-sequence B-type pulsators\thanks{This work resulted from the Honour's Bachelor project offered to the first author by Radboud University Nijmegen, by means of an internship in the asteroseismology group of KU\,Leuven.}}
\author{Rebecca Rehm\inst{\ref{KUL},\ref{Radboud}}
\and Joey S. G. Mombarg\inst{\ref{irap}, \ref{KUL}}
\and Conny Aerts\inst{\ref{KUL},\ref{Radboud},\ref{MPIA}}
\and Mathias Michielsen\inst{\ref{KUL}} 
\and Siemen Burssens\inst{\ref{KUL}} 
\and Richard~H.~D. Townsend\inst{\ref{Wisconsin}}
}
\institute{
Institute of Astronomy, KU Leuven, Celestijnenlaan 200D, B-3001 Leuven, Belgium \\ \email{rebecca-rehm2000@gmx.de} \label{KUL} 
\and Department of Astrophysics, IMAPP, Radboud University Nijmegen, PO Box 9010, 6500 GL Nijmegen, The Netherlands\label{Radboud}
\and
IRAP, Universit\'e de Toulouse, CNRS, UPS, CNES, 14 avenue \'Edouard Belin, F-31400 Toulouse, France \\ \email{jmombarg@irap.omp.eu} \label{irap}
\and Max Planck Institute for Astronomy, K\"onigstuhl 17, 69117 Heidelberg, Germany \label{MPIA}
\and
Department of Astronomy, University of Wisconsin-Madison, 475 Charter Street, Madison, WI, 53706, USA \label{Wisconsin}
}

\date{Received 15 February 2024 / Accepted 3 May 2024}

	\abstract 
	{Numerical computations of stellar oscillations for models representative of B-type stars predict fewer modes to be excited than observations reveal from modern space-based photometric data. One shortcoming of state-of-the-art evolution models of B-type stars that may cause a lack of excited modes is the absence of microscopic diffusion in most such models.}
	{We investigate whether the inclusion of microscopic diffusion in stellar models of B-type stars, notably radiative levitation experienced by isotopes, leads to extra mode driving by the opacity mechanism compared to the case of models that do not include microscopic diffusion.} 
	{We consider the case of slowly to moderately rotating stars and use non-rotating equilibrium models, while we account for (uniform) rotation in the computations of the pulsation frequencies. We calculate
1D stellar structure and evolution models with and without microscopic diffusion and examine the effect of radiative levitation on mode excitation, for both low-radial order pressure and gravity modes and for high-radial order gravity modes.
As is common practice in asteroseismology, rotation is included in the pulsation computations according to the mode's frequency regime. For modes having frequencies below twice the rotation frequency, that is, modes in the sub-inertial regime, we adopt the traditional approximation of rotation.  For modes in the super-inertial regime with frequency above twice the rotation frequency, rotation is treated perturbatively up to first order in the rotation. We consider macroscopic envelope mixing induced by internal gravity waves to compute the modes and study its effect on the surface abundances. }
	{We find systematically more modes to be excited for the stellar models including microscopic diffusion compared to those without it, in agreement with observational findings of pulsating B-type dwarfs. Furthermore, the models with microscopic diffusion predict that excited modes occur earlier on in the evolution compared to modes without it.
In order to maintain realistic surface abundances during the main sequence, we include macroscopic envelope mixing by internal gravity waves. Along with microscopic diffusion, such macroscopic envelope mixing ensures both more excited modes and surface abundances consistent with spectroscopic studies of B-type stars. } 
	{While radiative levitation has so far largely been neglected in stellar evolution computations of B-type stars for computational convenience, it impacts mode excitation predictions for stellar models of such stars.  We conclude that the process of radiative levitation is able to reduce the discrepancy between predicted and observed excited pulsation modes in B-type stars.  }

\keywords{Asteroseismology -- Atomic processes -- Stars: oscillations (including pulsations) -- Stars: interiors -- Stars: rotation -- Stars: evolution}

\titlerunning{The impact of radiative levitation on mode excitation of B-type pulsators}
\authorrunning{R. Rehm et al.}
\maketitle

\section{Introduction}

Non-radial oscillations of B-type stars are self-excited by the so-called $\kappa$-mechanism, converting thermal energy into mechanical energy. For the hot B-type pulsators this  mechanism 
operates in the partial ionization zone of the iron-group elements located in their outer envelope
at a temperature of about $T\sim 2 \times 10^{5} \, \si{\kelvin}$ \citep[e.g.][for a summary]{Moskalik1992,Dziembowski1993a,Dziembowski1993b,Gautschy1993, Pamyatnykh1999}. In this zone the opacity reaches a maximum due to increased contributions from  iron-like elements, notably iron and nickel. 
This zone acts as a heat-engine causing excitation of low-radial order pressure and gravity modes with periods of a few hours in the $\beta\,$Cep stars having an average mass around 9\,M$_\odot$ and effective temperature around 25\,000\,K \citep{Stankov2005}. The same mechanism excites high-radial order gravity modes with periods on the order of a day in the so-called Slowly-Pulsating B-type (SPB) stars having an average mass of some 4\,M$_\odot$  and effective temperature around 15\,000\,K \citep{Waelkens1991,DeCatAerts2002}. While the mode excitation in these two groups of pulsators is relatively well understood in terms of the $\kappa$-mechanism in rotating stars \citep{Szewczuk2017}, discrepancies occur between theoretical instability predictions and observed modes, across the entire range of rotation rates of the observed pulsators, from very slow to very fast rotators. 
More concretely, we observe several more modes in B-type pulsators than predicted by the theory of mode excitation. While it often concerns only one or a few missing modes 
\citep[e.g.][]{Pamyatnykh2004,Dziembowski2008,Moravveji2016b,Daszynska2017}, this mismatch between observed and predicted excited modes points to shortcomings in the theory of stellar structure. In particular,
at least one physical ingredient is missing or poorly described in current stellar structure models of B-type stars. Moreover, the lack of excited modes in the models compared to the observed modes occurs in both single and binary B-type pulsators. Hence close binarity cannot explain all of the excitation discrepancies. 

Differences between theoretical predictions of oscillation frequencies computed in the adiabatic approximation and a fully non-adiabatic description are small compared to observational uncertainties, even in the current era of high-precision space asteroseismology \citep{Aerts2018}. Given that adiabatic pulsation computations are much less CPU-demanding than non-adiabatic ones, forward asteroseismic modelling of individual pulsators is usually done in an adiabatic framework \citep{Aerts2003,Briquet2007,Dziembowski2008,Handler2009b,Aerts2011,Briquet2012,Aerts2019,Pedersen2021}.
This saves considerable CPU time for current-day grid-based approaches in forward modelling, 
which is a high-dimensional computationally intensive optimisation problem. 
A posteriori, instability tests are then often done for the few selected best forward models, 
revealing a lack of excited modes compared to observations in almost all modelled 
$\beta\,$Cep stars \citep{Pamyatnykh2004,Handler2009a,Daszynska2010,Daszynska2017} and SPB stars \citep{Moravveji2015,Moravveji2016,Szewczuk2018,Szewczuk2021,Szewczuk2022}. 
Discrepancies between mode excitation computations and observations now also occur 
for the large samples of OB-type pulsators discovered in the time-series photometry from NASA's TESS mission \citep[e.g.][]{Burssens2020} and from
Gaia Data Release 3, where many of such newly discovered pulsators occur outside the instability regions of B-type pulsators \citep{DeRidder2023,Aerts2023}.
We cannot help but conclude that evidence for missing input physics in stellar structure models is piling up from the viewpoint of mode instability computations. 

While modern space photometry showed that nonlinear mode coupling may also play a vital role in getting extra modes excited in large-amplitude $\beta\,$Cep \citep{Degroote2009} and SPB \citep{VanBeeck2021} stars, we focus here on the question how to excite more self-driven modes excited by the $\kappa$-mechanism. 
In order to excite more $\kappa-$driven modes in B-type pulsators, a higher opacity and hence higher abundances of the iron-like species in their partial ionisation layers are required.  Attempts to excite more modes by artificially increasing the opacity in the iron-opacity bump region 
\citep{Pamyatnykh2004,Salmon2012, Daszynska2017} or in the star as a whole 
\citep{Moravveji2016b} 
have been proven to be successful to excite the observed modes. However, an artificial local increase of opacity is not satisfactory from a physical point of view. A promising physical process for improving mode excitation is microscopic diffusion \citep{Michaud2004}. While microscopic diffusion has been included in models of A-type stars to explain surface abundances \citep{Deal2016} and in asteroseismic models of F-type gravity-mode pulsators \citep{Mombarg2020,Mombarg2022}, it was not yet considered as a global physical ingredient in structure models of B-type stars, mostly due to computational challenges
\citep{Bourge2006,Bourge2007}. 

In this paper we assess the potential of microscopic diffusion, including radiative levitation, to increase the number of excited modes in stellar models of B-type stars. We do so by focusing on two representative stellar models: a 4\Msun model typical of an SPB pulsator having high-radial order gravity (g) modes and a 9\Msun model typical of  a $\beta$~Cep star exhibiting low-radial order pressure (p) and g modes. For both models two evolutionary tracks are calculated, one with microscopic diffusion and one without it. For various stages along the main sequence we then compare the 
modes predicted to be excited by the $\kappa$-mechanism for the case with and without microscopic diffusion.

\section{Computational setup} \label{sect:}
\subsection{Frozen input physics of the stellar models}
We assume that any star represented within our grid of models rotates slow enough to ignore rotational deformation due to the centrifugal force. Moreover, mass loss is not taken into account. Hence we compute spherically symmetric stellar models in hydrostatic equilibrium\footnote{Our computational setup can be found at \url{https://doi.org/10.5281/zenodo.10992486}.}. 
These were computed using the stellar structure and evolution code \mesa \citep{Paxton2011, Paxton2013, Paxton2015, Paxton2018, Paxton2019, Jermyn2023}, version r23.05.1. 

The models adopt OP opacity tables \citep{Seaton2005} and the standard chemical mixture of OB stars in the solar neighbourhood based on \citet{Nieva2012}. Isotopes of Ne, Na, Al, Si, S, Ar, Ca, Cr, Mn, Fe, Ni, Cu are added to the basic reaction network in \mesa. Following the same procedure as \citet{Michielsen2021}, the initial chemical composition is set by choosing an initial metallicity. Our models are computed for $Z_{\mathrm{ini}}=0.014$, and the initial helium-mass fraction is determined based on an enrichment law of the form $Y_{\mathrm{ini}}=Y_{\mathrm{p}} + (\Delta Y/\Delta Z)Z_{\mathrm{ini}}$, where $Y_{\mathrm{p}}$ is the primordial helium-mass fraction and $(\Delta Y/\Delta Z)$ is the galactic enrichment ratio. We use $Y_{\mathrm{p}}=0.2465$ from \citet{Aver2013} and $(\Delta Y/\Delta Z)=2.1$. The galactic enrichment ratio is calculated so that the chemical mixture of OB stars in the solar neighbourhood from \citet{Nieva2012} is reproduced, which is adopted here to set the relative metal abundances of the models because most pulsating B-type stars adhere to this chemical mixture from the viewpoint of the measured surface abundances \citep{Gies1992,Morel2006,Niemczura2009,Gebruers2021}. Subsequently the initial hydrogen-mass fraction is calculated as $X_{\mathrm{ini}}=1-Y_{\mathrm{ini}}-Z_{\mathrm{ini}}$.

We model the mixing in the core convective regions using time-dependent convection \citep{Kuhfuss1986}. The boundary between the convective core and radiative envelope is determined by the Schwarzschild criterion. As convective boundary mixing (CBM) we include a step overshoot prescription with the radiative temperature gradient in the transition layer between the convective core and the radiative envelope. Thus the radial extent of the fully mixed region in the core is enlarged by
$\alpha_{\rm ov}\,H_P=0.2\,H_P$, where $H_P$ is the local pressure scale height.
An Eddington grey atmosphere is chosen as atmospheric boundary condition.

Since we are interested in the effects of microscopic diffusion, we calculate two models that differ only in that one is with microscopic diffusion (including radiative levitation) and the other without. The accelerations due to radiative levitation ($g_{{\rm rad}, i}$, with $i$ the species index) need to be calculated from monochromatic opacities. We relied on the OP data to do so \citep{Seaton2005}. In addition, the Rosseland mean opacity ($\kappa_{\rm R}$) depends on the local mixture, and since the process of microscopic diffusion changes the relative metal fractions, the opacity cannot be interpolated from precomputed tables for a specific mixture, but need to be computed from monochromatic opacities. This significantly increases the CPU time to compute stellar evolution models with microscopic diffusion. 
In order to deal with this high computational demand and to speed up the computations with radiative levitation, we used the routines based on \cite{Mombarg2022}, which are available in \mesa version r23.05.1 \citep{Jermyn2023}. In this method, the grids on which the \grad and \kappar are interpolated in temperature and density are recomputed if the maximum difference in the fractional abundances exceeds a certain threshold. 

To obtain zero-age main sequence models, the pre-main sequence computations are calculated with atomic diffusion already included, while overshooting is neglected. Lastly, we verified that the results presented in this paper are robust for the temporal resolution that we choose.  

\begin{figure}[h]
    \centering
    \includegraphics[width=\columnwidth]{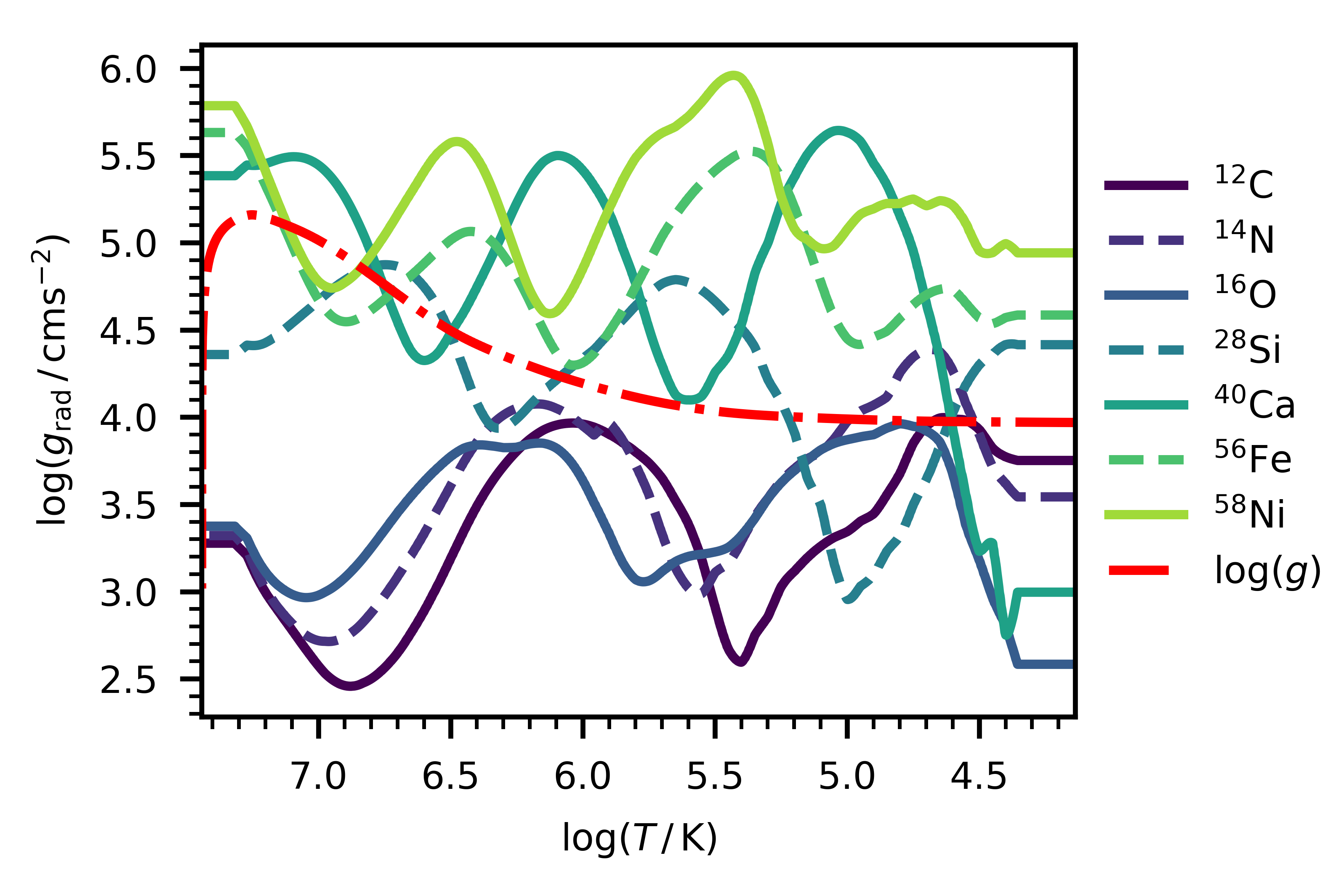}
    \caption{Gravitational acceleration (dashed-dotted red line) and acceleration due to radiative levitation for selected elements in the 4\Msun star ($Z_{\rm{ini}}=0.014$, $\log{D_{0}}=3$, $\alpha_{\rm ov}=0.2$, $\zeta=-0.5$) at $X_{\rm c}=0.36$, where the definitions of $D_0$ and $\zeta$ are given in Eq.~(\ref{eq:Dmix}).}
    \label{fig:grad_profile_M4.00_a0.5}
\end{figure}

\subsection{Macroscopic mixing in the envelope} \label{sec:IGW_mix}

\begin{figure*}
    \centering
    \includegraphics[width = 18cm]{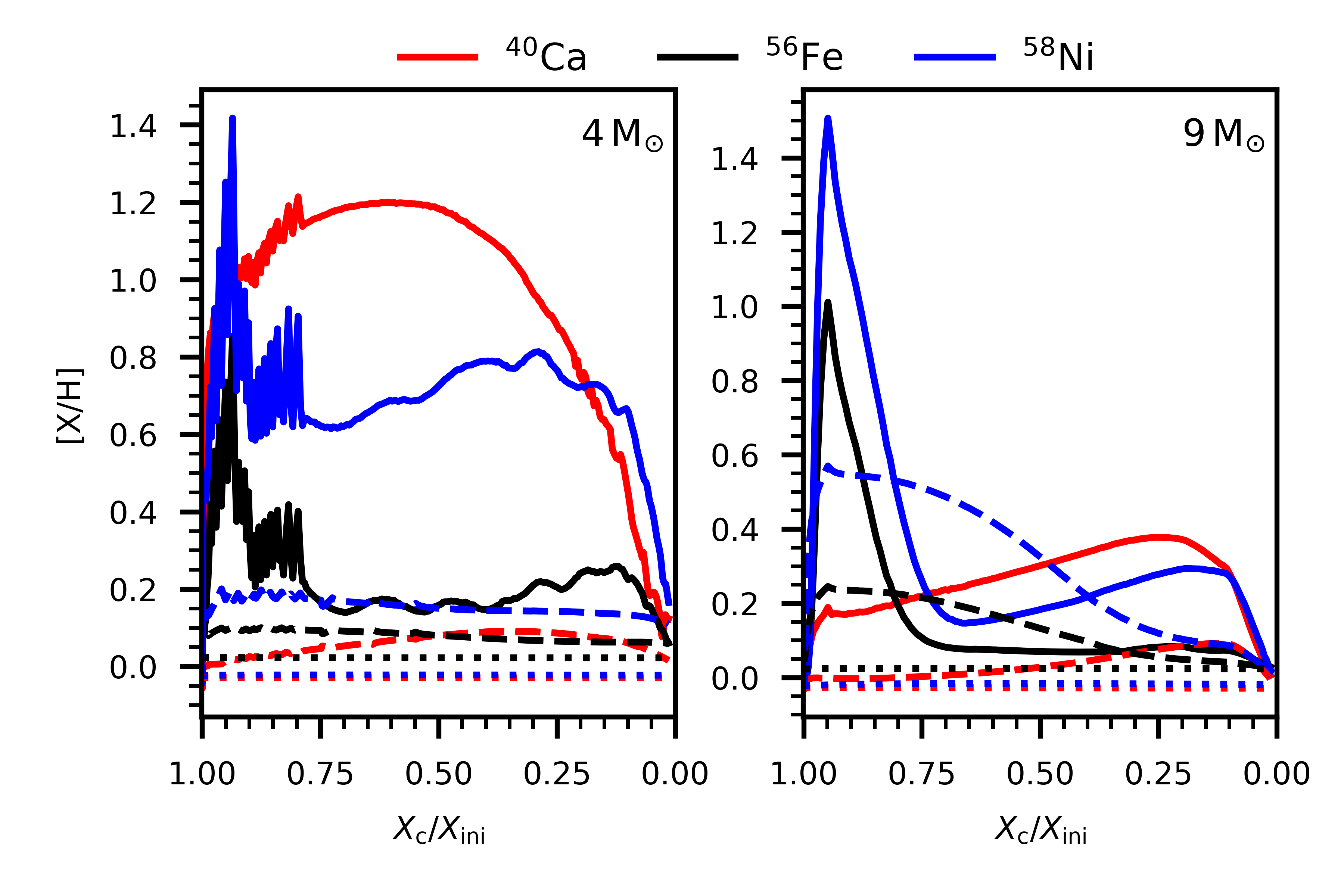}
    \caption{Surface abundances (${\rm [X/H]} = \log{({\rm X/H})} - \log{({\rm X/H})}_\odot$) with respect to the hydrogen surface abundance as a function of fraction of initial hydrogen burnt in the convective core. The surface abundances are scaled with the Solar composition as determined by \cite{Magg2022}. Solid lines correspond to $\zeta = 0$, dashed lines to $\zeta = -0.5$, and dotted lines to $\zeta = -1$. }
    \label{fig:surf_abun}
\end{figure*}

We do not consider rotational mixing in the radiative envelope since we focus on slow to modest rotators for this study.  
Fig.~\ref{fig:grad_profile_M4.00_a0.5} shows the resulting radiative acceleration for some metal isotopes, as well as the local gravitational acceleration, in the 4\Msun model roughly halfway through the core-hydrogen burning phase. Numerical stability requires for the outer most parts of the star to be treated as a single cell \citep{Mombarg2022}.  Hence, we choose to keep \grad constant over the outer most $10^{-9}$ part of the mass. In the locations where the radiative acceleration exceeds the gravitational one, the element species experiences a net migration towards the stellar surface. Hence, the surface abundances of $^{28}$Si, $^{56}$Fe, and $^{58}$Ni increase over time, while $^{1}$H is settling towards the core. This effect is even stronger in the 9\Msun case, as the radiative acceleration scales with a star's luminosity. This yields values of [Fe/H] that are unrealistically high compared to spectroscopic measurements of B-type stars \cite[e.g.][]{Gies1992,Morel2006,Niemczura2009,Gebruers2021}, as can be seen in Fig.~\ref{fig:surf_abun}. Since the efficiency of microscopic diffusion is computed from first principles \citep{Michaud2004}, the radiative envelopes of dwarfs must experience also a form of macroscopic mixing, partially counteracting the effects of microscopic diffusion. 
This is the reason why macroscopic envelope mixing is usually included as an extra ingredient when computing models of dwarfs with a convective surface layer.
This turbulent mixing may have different formulations but is chosen such as to obtain realistic surface abundances \citep[see e.g.][for further explanations and arguments]{Morel2002, Choi2016, Dotter2017}.

Models including two types of mixing, which are due to microscopic diffusion and macroscopic phenomena in the radiative envelope, show promising potential to explain the observed structure in the gravity mode period spacing patterns of pulsating F-type ($\gamma$~Doradus) stars  \citep{Mombarg2022}. Following the need of a radially-dependent chemical diffusion coefficient found from forward asteroseismic modelling of a sample of SPB stars by \citet{Pedersen2021} and our choice to consider slow to moderate rotators, we
assume a mixing profile as expected from internal gravity waves (IGWs) to be the dominant source of macroscopic mixing in the models. Hydrodynamical simulations show that the effective chemical diffusion coefficient from IGW mixing relates to the local density \citep{Rogers2017},
\begin{equation} \label{eq:Dmix}
    D_{\mathrm{IGW}}(r) = D_0 \left(\frac{\rho(r)}{\rho(r_0)}  \right)^{\zeta}, 
\end{equation}
where $\rho(r_0)$ is the density and $D_0$ the level of mixing at the point of transition between CBM and IGW mixing. The exponent $\zeta$ was found to have a value between $-1$ and $-0.5$ from the simulations \citep{Rogers2017}. The parametrization described by Eq.~(\ref{eq:Dmix}) is a simplification of the expression for the predicted wave amplitude from the linear theory of waves \citep{Ratnasingam2019}. The study by \citet{Varghese2023} presents predicted mixing profiles resulting from hydrodynamical simulations of IGW excitation in stars between 3 and 20\Msun, covering models from close to the zero-age main sequence up to close to the terminal-age main sequence. Their results show that the radial dependence of the chemical diffusion coefficient varies significantly during the main sequence. They also find that the predicted wave amplitude from linear theory overpredicts the wave amplitude for stars nearing the end of core-hydrogen burning. In anticipation of a time-dependent implementation of IGW mixing in \mesa, we leave $\zeta$ as a free parameter and choose its value such that additional modes are excited in the models with microscopic diffusion, while still maintaining realistic abundances at the stellar surface.

\begin{figure*}
    \centering
    \includegraphics[width = 0.9\textwidth]{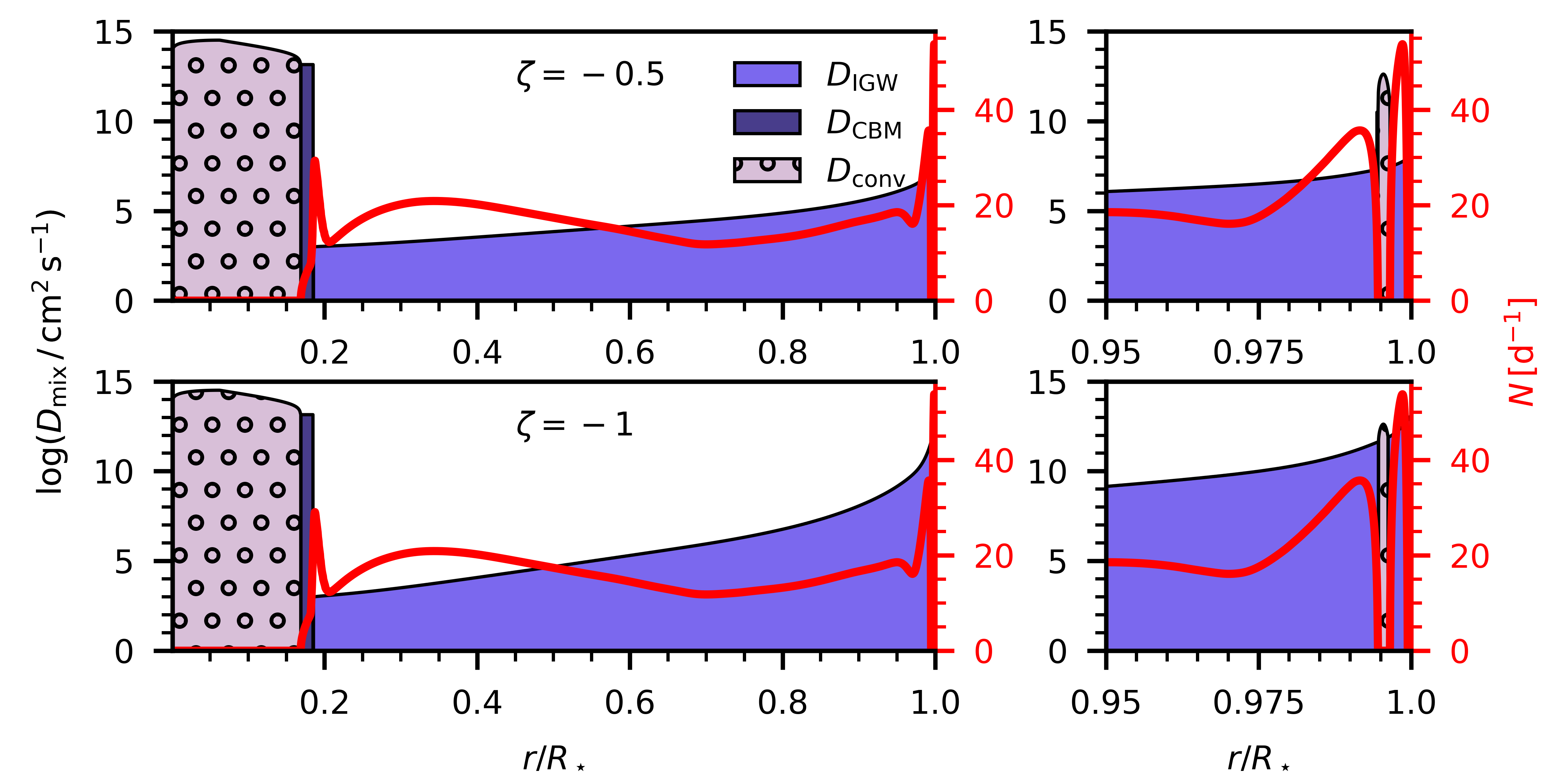}
    \caption{Variation of the chemical diffusion coefficient as a function of the radial coordinate throughout the star ($4\,{\rm M}_\odot$, $Z_{\rm{ini}}=0.014$, $\log{D_{0}}=3$, $X_{\rm c} = 0.66$). Different types of mixing regions are indicated. The top panel corresponds to setting $\zeta = -0.5$ in Eq.~(\ref{eq:Dmix}), the bottom panel to setting $\zeta = -1$. The graphs in red show the corresponding profile of the Brunt-V\"ais\"al\"a frequency ($N$). The right column shows zoom-ins of the outermost five per cent of the radius.  }
    \label{fig:Dmix}
\end{figure*}

Fig.~\ref{fig:Dmix} shows the difference in the resulting chemical diffusion coefficient in the radiative envelope depending on the density-dependence of the IGW mixing. In the outer five per cent of the star's radius, where the partial ionization zones are located, the diffusion coefficient varies two-to-three orders of magnitude, depending on the choice of $\zeta$. 
The resulting evolution of the surface abundances of isotopes $^{40}$Ca, $^{56}$Fe, and $^{58}$Ni are shown in Fig.~\ref{fig:surf_abun} for $\zeta \in [-1, -0.5, 0]$. 
The small-scale variations are attributed to numerical noise, but the global trend is physical. As can been seen in the model with $\zeta = -1$, the macroscopic mixing near the surface is too efficient to see any variations as a result of microscopic diffusion for this value of the exponent. 
The summary plot shown in Fig.~19 of \citet[][see also references therein]{Gebruers2021} shows that the measured $^{56}$Fe surface abundances of most bright B-type stars in our neighbourhood range from ${\rm [X/H] = -0.6}$ to $+0.15$. A power law with a coefficient of $\zeta = -0.5$ is thus compatible 
with observed $^{56}$Fe surface abundances of B-type pulsators throughout the main sequence for both 4 and 9\Msun. For this reason, we use this value in the envelope mixing profile. In Sect.~\ref{sect:discussion}, we further discuss the implications of a stronger density-dependence for the efficiency of IGW mixing on mode excitation. 

\subsection{Pulsation computations}

In order to compute the eigenmodes of a star, one needs to perturb its equilibrium model and solve an eigenvalue problem described by the perturbed equations of stellar structure \citep{Aerts2010}. Depending on the rotation of the star, the Coriolis acceleration can or cannot be treated as a small perturbation. The key quantity to determine the best approximation for the computation of a particular oscillation mode is its spin parameter, $s\equiv 2\Omega/\omega$, where $\Omega$ is the rotation frequency of the star and $\omega$ is the mode's frequency in a corotating frame. When $s\leq 1$ 
the modes are in the super-inertial regime and one may adopt a perturbative approach in $\Omega$ to compute the modes. For $s>1$, on the other hand, the modes are sub-inertial and one has to include the Coriolis force in a non-perturbative approach \cite[e.g. see][for an extensive discussion of asteroseismic applications in these two frequency regimes]{Aerts2024}. 

In this work, we compute linear non-adiabatic pulsation modes of the \mesa models with the stellar pulsation code \gyre \citep{Townsend2013,Townsend2018,Goldstein2020}, version 7.0. 
Almost all of the g~modes in B-type stars occur in the sub-inertial regime \citep{Aertsetal2021}. In such a case, the so-called Traditional Approximation of Rotation (TAR) offers a suitable approach for the mode computations \citep{Townsend2003}, so we use \gyre in this setting. We restrict ourselves to the computation of prograde dipole modes ($\ell =1, m = 1$) with radial orders between $n_{\rm pg} \in [-80, -1]$, given that these are the modes occurring most frequently in rotating SPB stars \citep{Pedersen2021}.  We
compute the eigenmodes from the pulsation equations in the TAR as an adequate approach for SPB pulsators \citep[e.g.][]{Aerts2021}. We adopt a uniform rotation with a frequency representing a moderate rotator, namely 20 per~cent of the Roche critical rotation frequency. 
The frequency scan range is set according to the buoyancy travel time of the model \citep[see][for a definition]{Aerts2021} to ensure that no radial orders are skipped in the \gyre computations. Finally, we rely on the outer-boundary conditions as described by \cite{Unno1989}.

In slow and moderately rotating $\beta\,$Cep stars, low-radial order p and g~modes are the dominantly observed modes \citep[][among many other examples]{Aerts2003,Briquet2007,Dziembowski2008,Handler2009a,Handler2009b,Aerts2011,Briquet2012,Daszynska2017,Cotton2022,Burssens2023}. 
The p modes occur in the super-inertial regime and as such the Coriolis acceleration can be treated perturbatively in $\Omega$. We restrict to the computation of  zonal p-mode frequencies for dipole ($\ell =1, m = 0$) and quadrupole ($\ell =2, m = 0$) modes for the 9\Msun model.
These modes are representative of the central components of most observed rotationally split multiplets in $\beta\,$Cep stars that got scrutinised by asteroseismic modelling in the papers referenced in this paragraph.

\section{Predicted mode excitation} \label{sect:resultsg}
The excitation of modes in SPB and $\beta\,$Cep stars is driven by the $\kappa$-mechanism.
Whether a mode is excited or not is determined by the sign of the imaginary part of the non-adiabatic mode frequency. For excited modes the imaginary part of the frequency ($\omega$) is positive, while it is negative for non-excited modes. Inspection of the so-called work function \citep[see ][their Eq.~(1) and its discussion]{Pamyatnykh1999} can indicate which parts of a star contribute toward excitation or damping of a mode; overall, ${\rm Im}(\omega) > 0$ occurs when the net excitation exceeds the net damping, and vice versa. The expression for the work integral involves the perturbation of the luminosity, in which the Lagrangian perturbation of the opacity, $\delta\kappa/\kappa$, plays the key role.

The opacity in the envelopes of main sequence stars is dominated by radiative processes. One thus considers the Rosseland mean opacity, which 
depends on the local fractional abundances of the individual species, $f_k$, in the following way,
\begin{equation} \label{eq:kap}
    \kappa_{\rm R} = \frac{1}{\mu}\left( \int_{0}^{\infty} \frac{{\rm d}u}{\sum_k f_k(r)\sigma_k(u)} \right)^{-1},
\end{equation}
where $\mu$ is the mean molecular weight, and $\sigma_k$ the monochromatic cross sections. The latter is defined as a function of $u = h\nu/k_{\rm B}T$, where $h$ is Planck's constant, $\nu$ the photon frequency, $k_{\rm B}$ Boltzmann's constant, and $T$ the temperature.

As explained in detail in \citet{Pamyatnykh1999}, to which we refer for details, the opacity derivatives with respect to temperature and density determine whether a mode is excited or not, along with the conditions that the period of the mode must be comparable to the local thermal timescale and that the amplitude of the mode must be large in the driving region. Any physical ingredient that changes the opacity, such as microscopic diffusion, may affect the mode excitation. 
Previous studies that have attempted to resolve the lack of excited modes have increased the contribution of iron and nickel to the Rosseland mean opacity in an ad-hoc way. In terms of Eq.~(\ref{eq:kap}), this means multiplying $f_k\,\sigma_k$ for these two species with some factor. As follows from this integral equation, increasing the fractional abundances locally is indeed sufficient to change the local opacity and its derivatives. The stability of a mode depends on the radial gradient of the following two derivatives
with respect to the density and temperature,
\begin{equation}
    \kappa_{\rho} = \frac{\partial \ln \kappa_{\rm R}}{\partial \ln \rho},~\kappa_{T} = \frac{\partial \ln \kappa_{\rm R}}{\partial \ln T},
\end{equation}
respectively. Mode excitation may change as well and it could follow naturally from the accumulation of heavy elements around the iron-opacity bump. 

\begin{figure}
    \centering
    \includegraphics[width = \columnwidth]{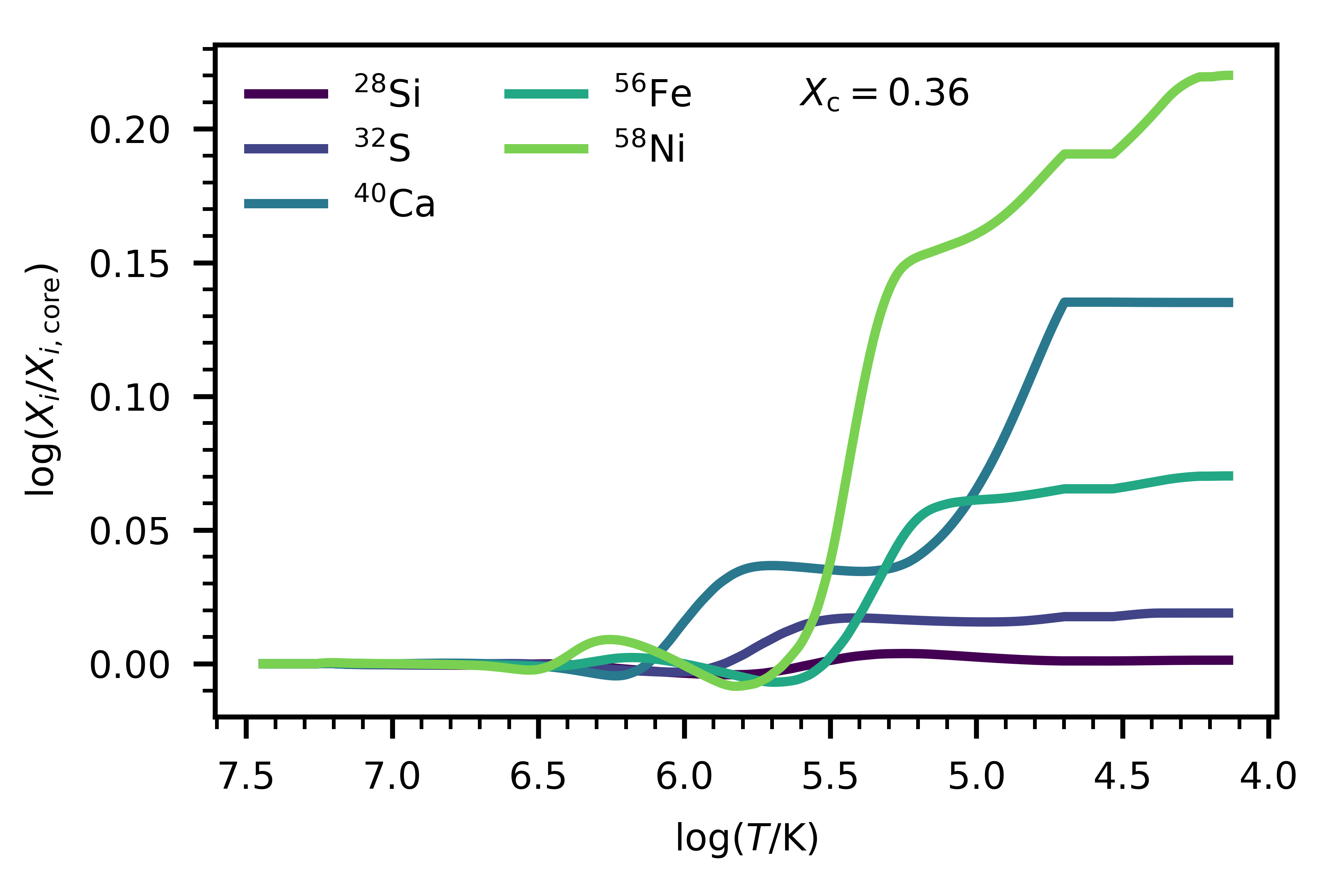}
    \caption{Logarithmic fraction of the local fractional abundance compared to the value in the convective core. This plot shows a model for a 4\Msun star at $X_{\rm c} = 0.36$ ($Z_{\rm{ini}}=0.014$, $\log{D_{0}}=3$, $\alpha_{\rm ov}=0.2$, $\zeta=-0.5$). }
    \label{fig:mass_frac}
\end{figure}

Fig.~\ref{fig:mass_frac} illustrates the increase of the fractional abundances of heavy elements towards the surface as a result of radiative levitation (chemical profiles at different ages are shown in Appendix~\ref{ap:chem_prof}). From this figure it can be seen that in the case of the model with 4\Msun, the $^{56}$Fe and $^{58}$Ni abundances are increased by roughly 20 and 50 per cent, respectively, around the iron-opacity bump. The resulting effect of the increased heavy-element abundances towards the surface is illustrated in Fig.~\ref{fig:opacity_profile_M4.00_a0.5}, and in Fig.~\ref{fig:opacity_profile_M9.00_a0.5} in Appendix~\ref{ap:opacity}. The Rosseland mean opacity, and thus its derivatives for models with microscopic diffusion is larger around $\log(T/{\rm K}) = 5.2$, where the g~modes are driven, compared to the values for the models without diffusion. The difference  decreases as the effects of IGW mixing become more dominant.
 
\begin{figure}
    \centering
    \includegraphics[width=\columnwidth]{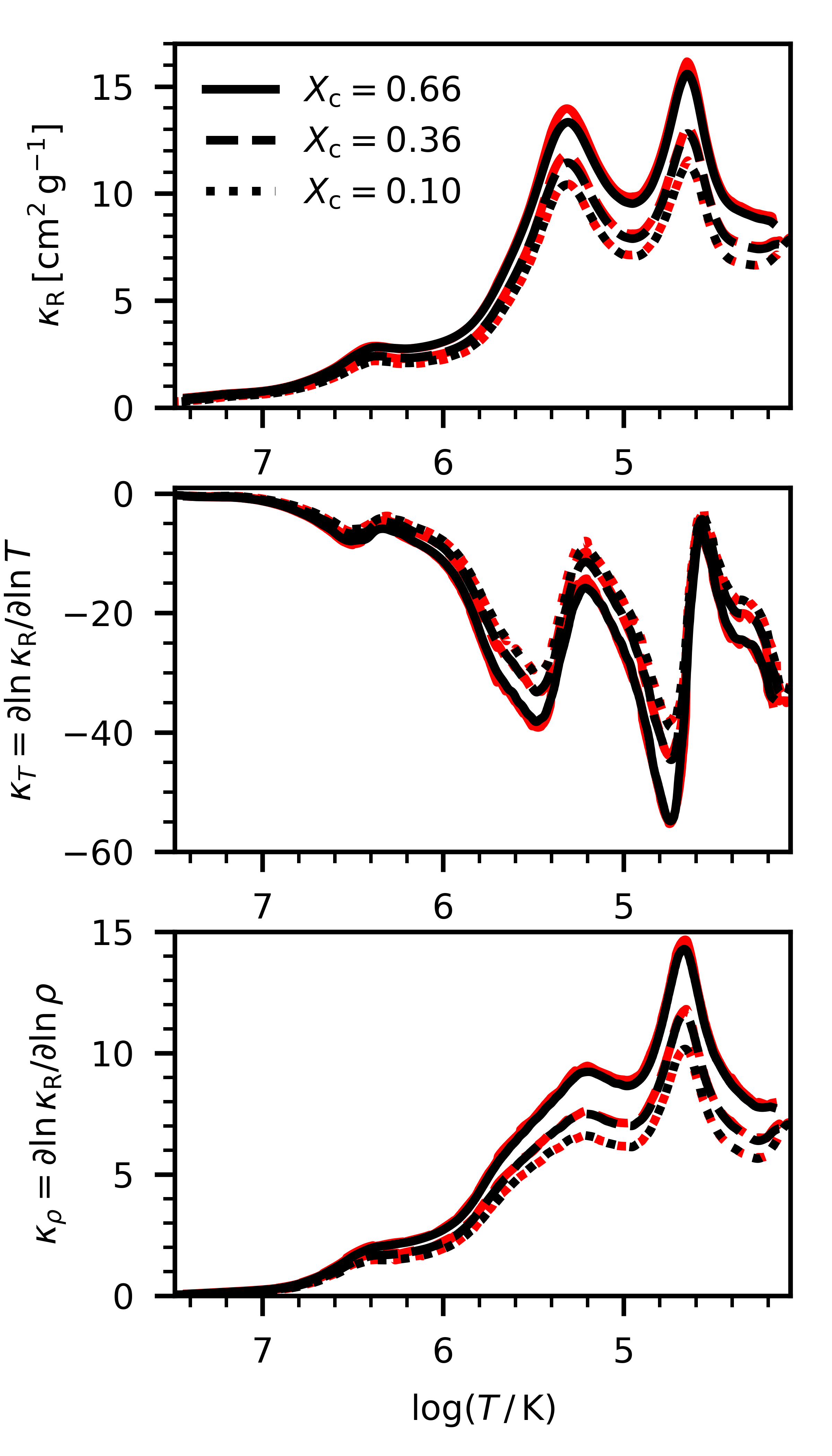}
    \caption{Rosseland mean opacity (top panel), its logarithmic derivative with respect to temperature (middle panel), and its logarithmic derivative with respect to density (bottom panel). Models are shown with diffusion (red) and without diffusion (black) of the 4\Msun model ($Z_{\rm{ini}}=0.014$, $\log{D_{0}}=3$, $\alpha_{\rm ov}=0.2$, $\zeta=-0.5$) at selected $X_{\rm c}$ values.}
    \label{fig:opacity_profile_M4.00_a0.5}
\end{figure}

\begin{figure*}[ht!] 
\centering
\begin{subfigure}{.5\textwidth}
  \centering
  \includegraphics[width=\linewidth]{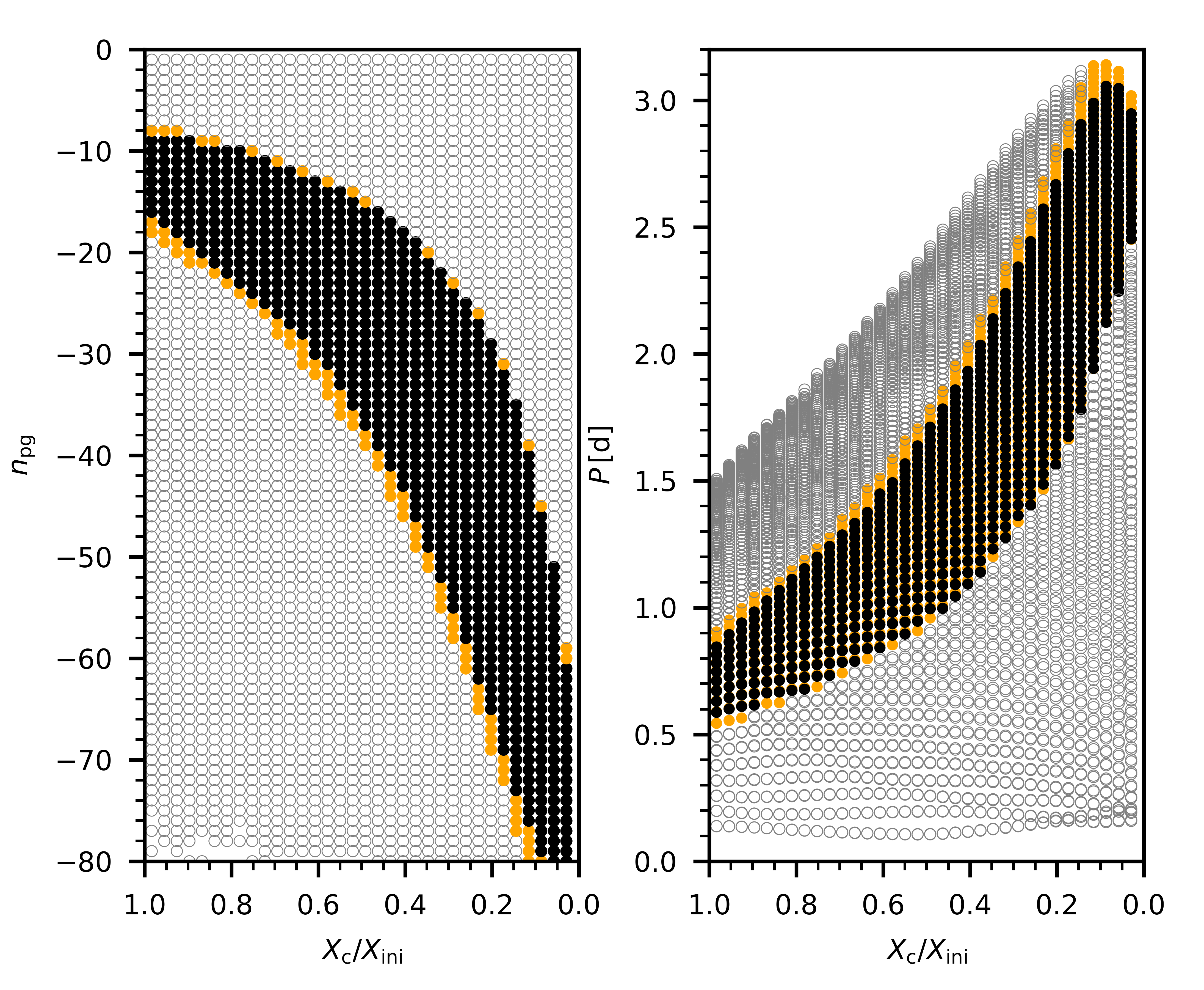}
\end{subfigure}%
\begin{subfigure}{.5\textwidth}
  \centering
  \includegraphics[width=\linewidth]{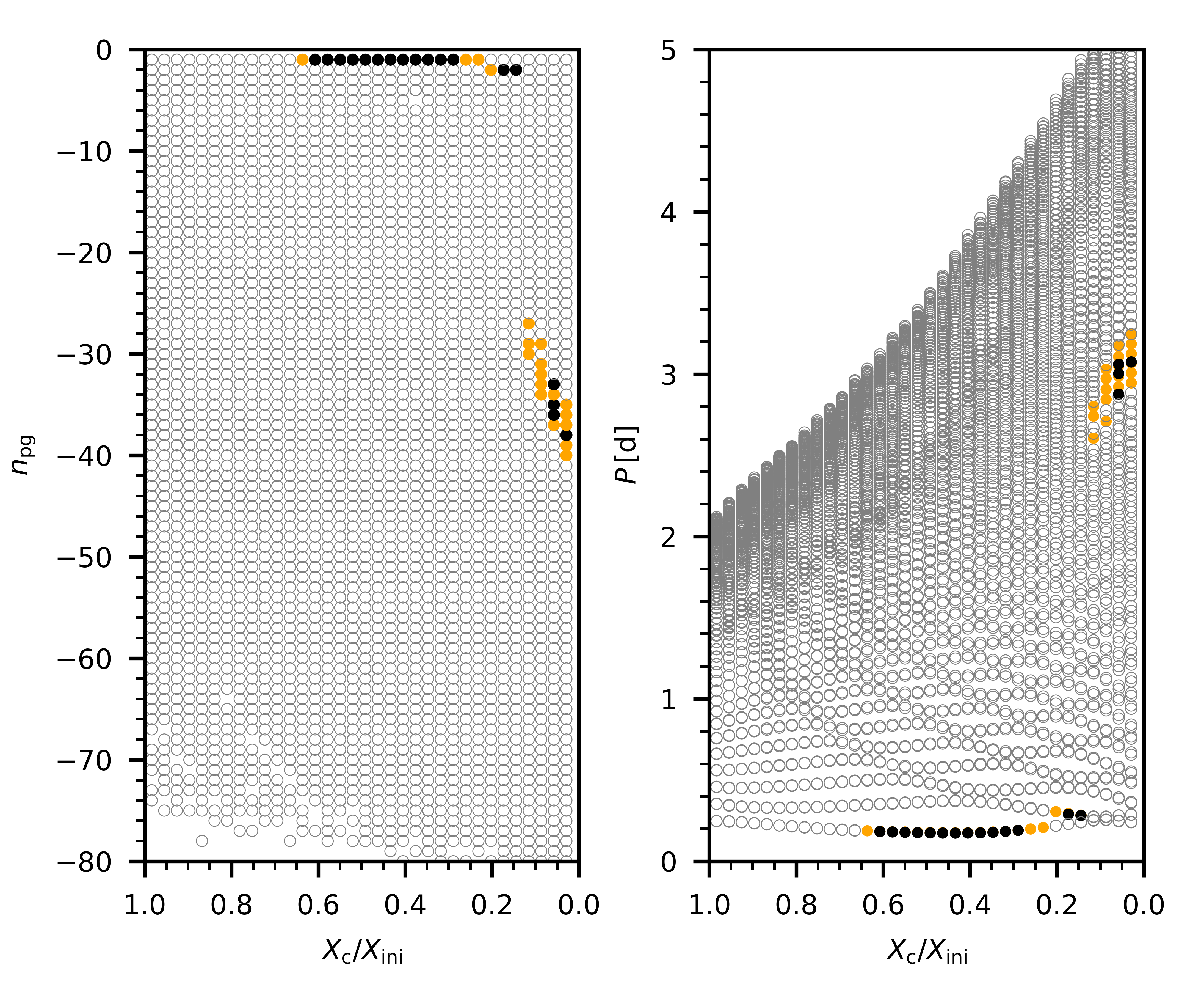}
\end{subfigure}
\caption{Radial orders and mode periods of the ($\ell=1, m=1$) g modes of a 4\Msun (two left-most panels) and a 9\Msun (two right-most panels) model, both having $Z_{\rm{ini}}=0.014$, $\log{D_{0}}=3$, $\alpha_{\rm ov}=0.2$, $\zeta=-0.5$, as a function of the central hydrogen mass fraction with respect to the initial value. A (uniform) rotation frequency equal to 20 per~cent of the critical Roche rotation frequency was taken for the computations of the pulsations. Filled dotes correspond to modes only excited in the models with microscopic diffusion (orange), and modes excited in both cases (black). Unfilled dots represent non-excited modes. In the panels with the mode period on the ordinate, the black dots indicate the model without microscopic diffusion. We note that the shifts in the mode periods between models with and without microscopic diffusion is small on the scale of this plot.
}
\label{fig:mode_excitation}
\end{figure*}

The results of our non-adiabatic dipole mode calculations are illustrated in Fig.~\ref{fig:mode_excitation}, where we show which radial orders are predicted to be excited as a function of the fraction of hydrogen left in the convective core. The orange dots show that the models with microscopic diffusion have several more modes excited compared to the models without it, particularly at the long period end of the mode spectrum. 
In their forward modelling of the slowly rotating SPB star KIC\,10526294, \citet[][their Models\,4 and 8]{Moravveji2015} required a twice as high metallicity as observed in spectroscopy.
This high metallicity was needed to get the 6 modes with longest period out of the 19 observed modes excited for models without atomic diffusion.  Our result in Fig.~\ref{fig:mode_excitation} illustrate that radiative levitation results in several more excited modes in a natural way, particularly in the period regime above 1.3\,d or longer.
Moreover, the best forward asteroseismic models for the moderately rotating 3.3\,M$_\odot$ SPB star KIC\,7760680 
 also excite about five dipole prograde modes too few for the period regime between 1.4\,d and 1.5\,d \citep{Moravveji2016}. Repeating the forward modelling using models with radiative levitation 
 for both SPB stars KIC\,10526294 and KIC\,7760680 might solve (part of) the mode excitation problem for these stars, but is far beyond the scope of the current paper.

In addition, for the 9\Msun model, we predict several dipole g~modes to be excited with periods between 2.4 and about 3\,d. This is a region of excitation where two g~modes were observed already two decades ago in the archetypical $\beta\,$Cep star $\nu\,$Eri, while instability computations did not predict any modes in models for a 9\,M$_\odot$ $\beta\,$Cep star at that time \citep{Handler2004}. More recent BRITE space photometry revealed 
a total of 7 g modes being observed in $\nu\,$Eri. The modelling by \citet{Daszynska2017} confirms the need for higher opacities to excite all these observed g modes. Although detailed forward modelling of $\nu\,$Eri is again far beyond the scope of our current paper, we conclude that our theoretical
results based on models including radiative levitation are in general agreement with the observed g modes of 
this star without the need for an extra increase in the iron and nickel opacities. A similar conclusion holds for the 4 observed g~modes detected in the 9\,M$_\odot$ 
hybrid 
$\beta\,$Cep/SPB star $\gamma\,$Peg by \citet{Handler2009b}.

Our results shown in Fig.~\ref{fig:mode_excitation} also reveal excited g modes to occur earlier in the main-sequence phase in the case of the 9\Msun model when microscopic diffusion is taken into account compared to the 
case where it is not included in the models.
Therefore, we conclude that the inclusion of microscopic diffusion in the stellar models not only increases the number of predicted excited g modes, but may also alter the shape of the SPB instability region.

\begin{figure}[t!]
\centering
\begin{subfigure}{.5\textwidth}
  \centering
  \includegraphics[width=\columnwidth]{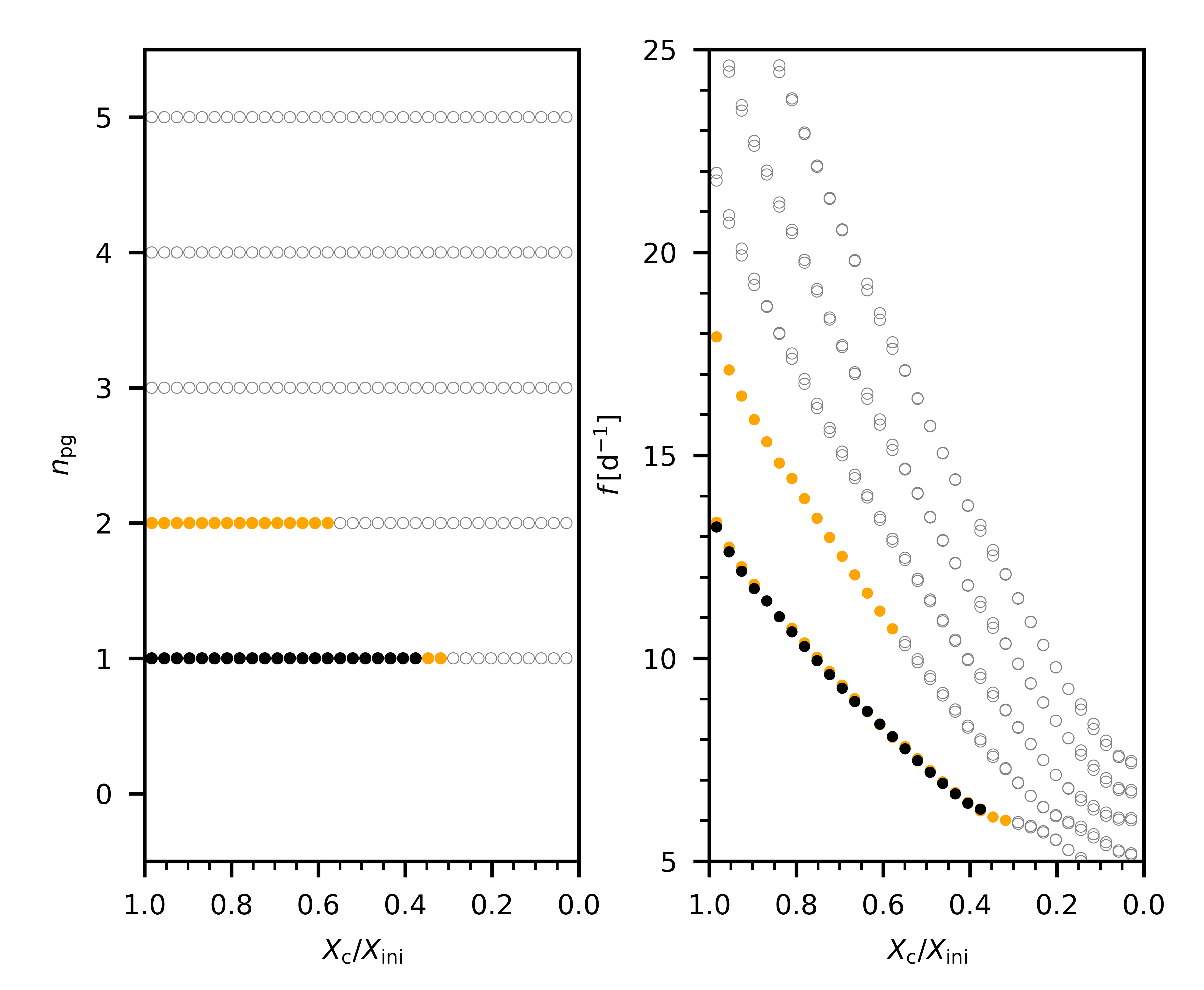}
  \caption{$\ell=1, m=0, \Omega = 0.2\,\Omega_{\rm crit}$}  
\end{subfigure}
\begin{subfigure}{.5\textwidth}
  \centering
  \includegraphics[width=\linewidth]{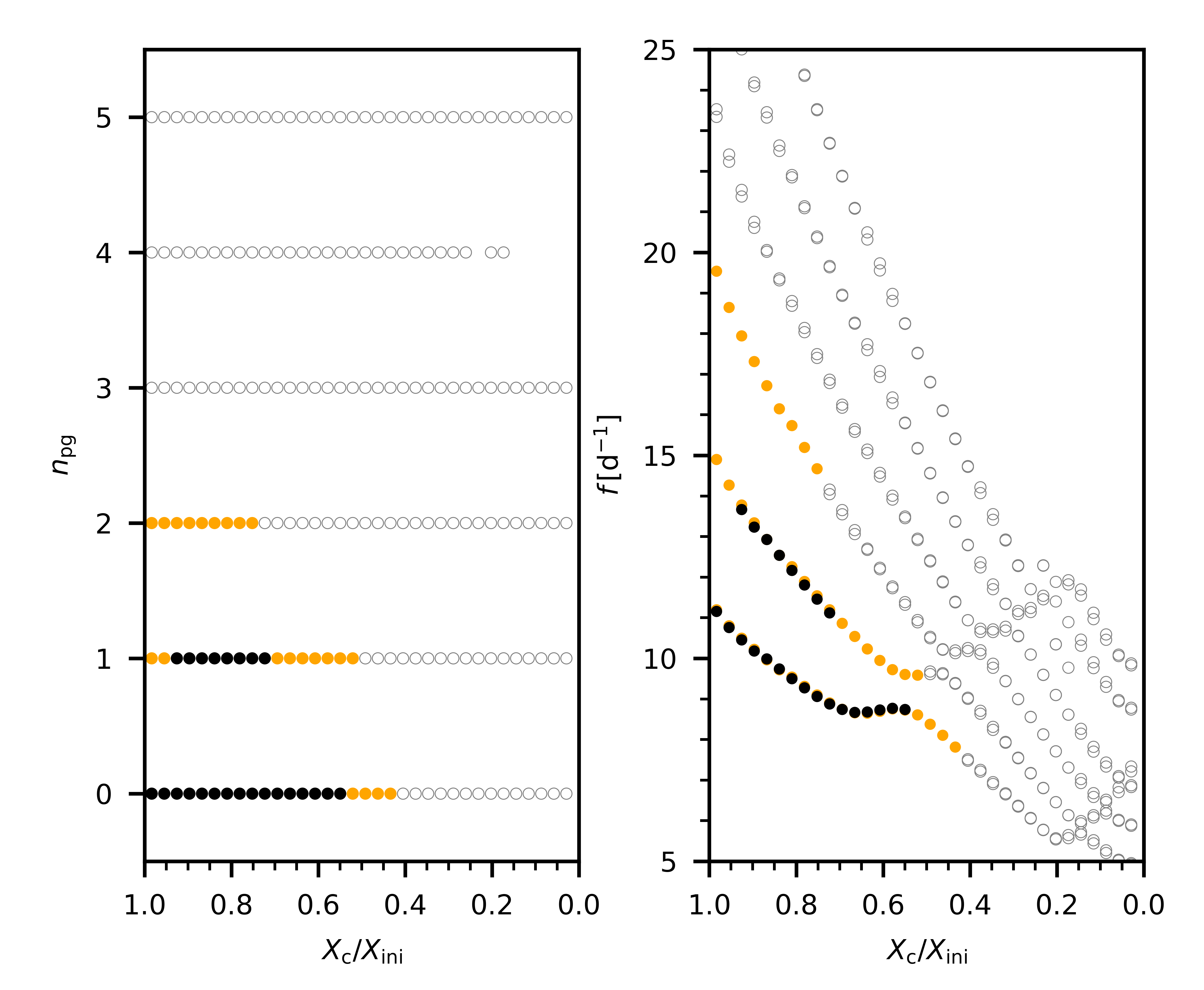}
  \caption{$\ell=2, m=0, \Omega = 0.2\,\Omega_{\rm crit}$ }
\end{subfigure}
    \caption{Same as Fig.\,\ref{fig:mode_excitation}, but for the radial orders 
of the p modes of the 9\Msun model.
}

\label{fig:mode_excitation_pmodes}
\end{figure}

For the 9\Msun model, we also perform non-adiabatic calculations for the low-radial order dipole and quadrupole zonal p~modes ($\ell=1, m=0$ and $\ell=2, m=0$).  The results are shown in Fig.~\ref{fig:mode_excitation_pmodes}. We find that the model with microscopic diffusion also predicts modes of radial order $n_{\rm pg} = 2$ to be excited, in addition to the $n_{\rm pg} = 1$ modes excited in both cases. Morevover, the $n_{\rm pg} = 1$ modes are excited for a larger fraction of the main sequence in the models with microscopic diffusion. Our findings may therefore offer a more natural explanation of the observed p modes in some $\beta\,$Cep stars without the need of higher (OP) opacities adopted in the models and instability regions computed by \citet{Moravveji2016b}.

\section{Influence of the level of envelope mixing} \label{sect:discussion}
\begin{figure}
\centering
\begin{subfigure}{.5\textwidth}
  \centering
  \includegraphics[width=\columnwidth]{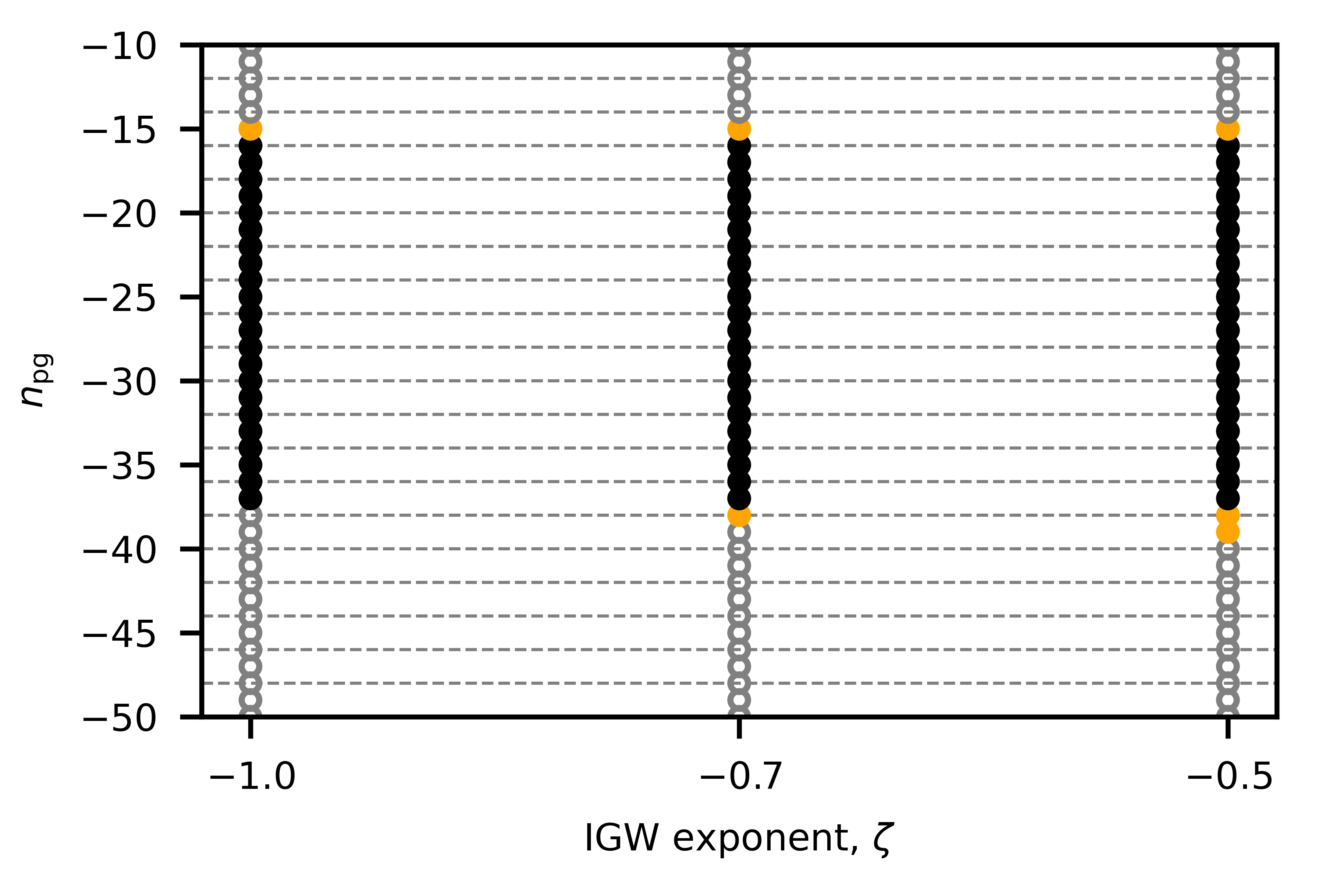}
  \caption{4\Msun, $X_{\rm c}=0.34$}  
\end{subfigure}
\begin{subfigure}{.5\textwidth}
  \centering
  \includegraphics[width=\linewidth]{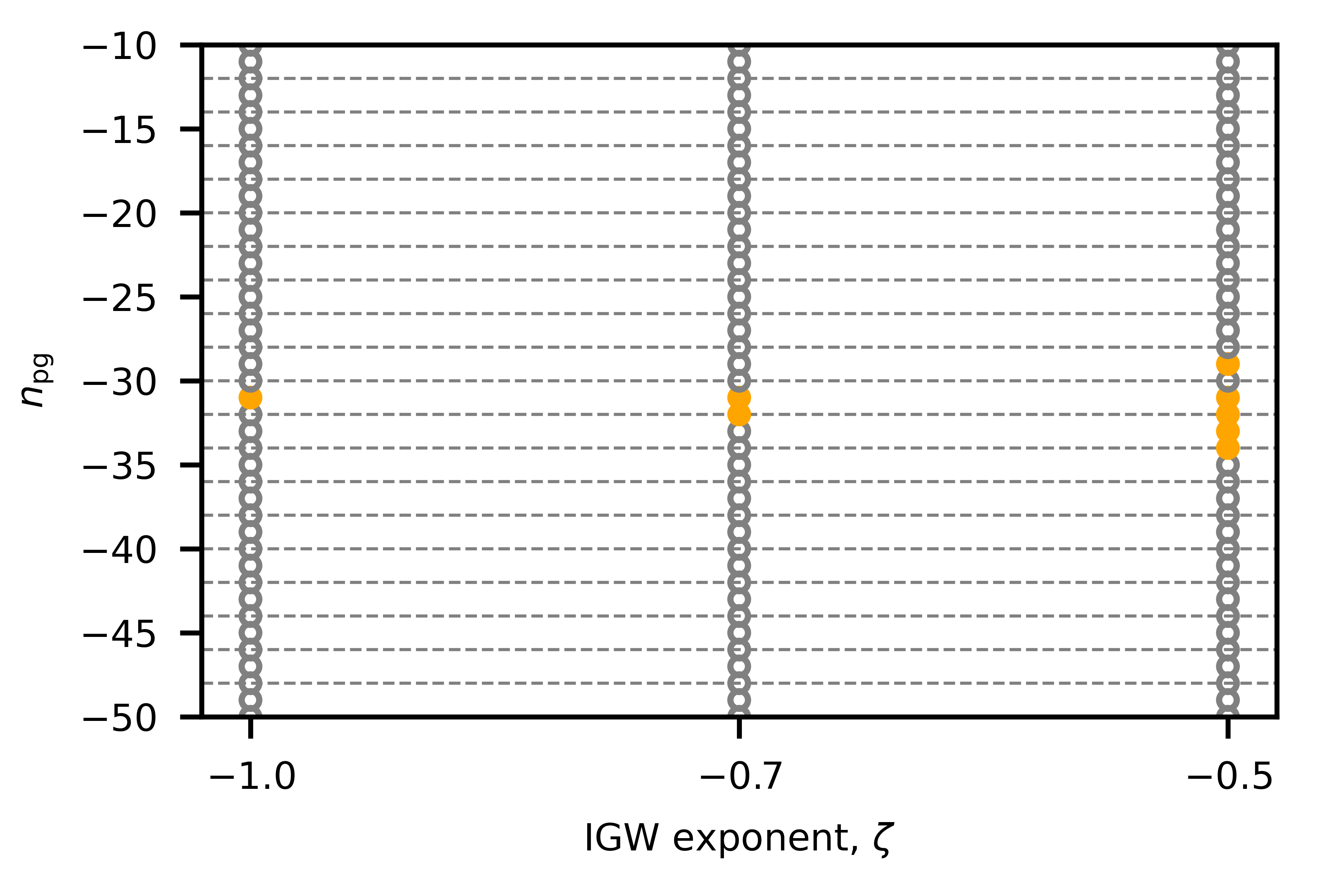}
  \caption{9\Msun, $X_{\rm c}=0.06$ }
\end{subfigure}
    \caption{Excited radial orders for models with IGW mixing ($Z_{\rm{ini}}=0.014$, $\log{D_{0}}=3$, $\alpha_{\rm ov}=0.2$, no rotation) with three different exponents $\zeta$ describing the density dependence (see Eq.~(\ref{eq:Dmix})). Filled black dots are excited in both models with and without microscopic diffusion, filled orange dots are excited in models with microscopic diffusion only and unfilled dots are non-excited modes.}

\label{fig:effect_IGW_exp_mode_excitation}
\end{figure}

In the stellar models discussed so far, we have limited ourselves to a fixed 
exponent of $\zeta=-0.5$
to describe the chemical diffusion coefficient from IGW mixing throughout the star. In Fig.~\ref{fig:effect_IGW_exp_mode_excitation}, we show the influence of the choice of $\zeta$ on the excited modes, using meaningful values in line with the hydrodynamical simulations by \citet{Varghese2023}. The figure shows the predicted excited g~modes for a 4\Msun model roughly halfway through the core-hydrogen burning phase, and a 9\Msun model close to the end of core-hydrogen burning, both for different choices of $\zeta$. We find that for $\zeta = -1$, the macroscopic mixing around the iron-opacity bump happens on a time scale that is short enough to prevent any significant build-up of iron and nickel, thus yielding no additional excited modes compared to the case without microscopic diffusion. 

For the case of $\zeta = -0.5$, Fig.~\ref{fig:effect_IGW_exp_mode_excitation} reveals extra g~modes to be excited at both ends of the radial order regime 
for models with microscopic diffusion compared to those without it.
This result illustrates that non-adiabatic mode excitation calculations have the potential to place tight constraints on the efficiency of IGW mixing in the outer envelope. However, the number of additionally excited g~modes is relatively small, and precise mass and age measurements are required. A procedure as in \citet{Moravveji2016, Michielsen2023} would be a good route, where forward modelling done from adiabatic calculations is followed by non-adiabatic computations for the (few) best resulting models, in order to tweak aspects of the mode excitation and level and kind of envelope mixing to higher precision.

Besides our choice of parameters to describe IGW mixing, the formalism as presented in Eq.~(\ref{eq:Dmix}) is also a simplification of reality. While hydrodynamical simulations carried out by \cite{Varghese2023} have shown that the scaling we assume in this work is a fairly accurate approximation, the IGWs are expected to experience additional damping as they pass the turning point marking their propagation cavity. The location of this turning point moves radially inwards throughout the main sequence for B-type stars \citep{Vanon2023} and is also mass dependent \citep{Rathish2023}.
The level of IGW mixing in the outer envelope of (slowly-rotating) B-type stars is expected to be orders of magnitude lower in more evolved stars. This implies that the 
difference in the local opacity between models with and without microscopic diffusion, as those shown in Figs.~\ref{fig:opacity_profile_M4.00_a0.5} and \ref{fig:opacity_profile_M9.00_a0.5} would be more pronounced towards the end of the main sequence.
Future improvements in the description of 1D mixing caused by IGWs as used here, where we relied on the time-independent formulation implemented by \cite{Pedersen2018}, will help to better quantify the evolutionary aspects of the mixing. This can be tackled by 
including time-dependent $D_{\rm IGW}$.

Given the computational demands of stellar evolution models with radiative levitation, we considered only one initial uniform composition according to observations of OB stars in the solar neighbourhood by \cite{Nieva2012}, and one initial metallicity of $Z_{\rm ini} = 0.014$. It is obvious that the initial chemical composition and metal mixture of the star is of major importance to assess the effects of microscopic diffusion for asteroseismology, notably the power of levitated iron and nickel isotopes entering the mode excitation layer near $2\times\,10^5$\,K.  When performing non-adiabatic asteroseismic modelling of concrete B-type stars, one has to compute dedicated models with their initial metal fractions and chemical mixture according to the measured surface abundances of the present-day star, keeping in mind its (often large) uncertainties.

\section{Conclusion}

Previous studies have attempted to resolve the discrepancy between observed excited pulsation modes in $\beta$~Cep and SPB stars and predictions from the theory of mode excitation by artificially increasing the opacities of iron and nickel in the driving zone \citep{Pamyatnykh2004, Moravveji2016b,Daszynska2017}. Such attempts show that ad-hoc increased opacities indeed result in more unstable modes. In this paper, we took an approach of including additional physics, derived from first principles, to the problem of observed excited modes missing in the models by making use of stellar models incorporating microscopic atomic diffusion along the evolutionary track of the star, including radiative levitation. We have shown that radiative levitation causes heavy elements to accumulate in the layers of the iron-opacity bump, where pulsation modes are driven in main-sequence B-type stars. We then computed non-adiabatic mode frequencies with \gyre to investigate which modes are predicted to be excited, using a 4\Msun and a 9\Msun stellar structure model. In both cases, we found that models with microscopic diffusion have extra excited modes compared to  models with the same input physics except for the microscopic diffusion. Furthermore, in case of the 9\Msun model, we also found high-radial order g~modes to be excited earlier in the main sequence phase compared to models without atomic diffusion.

Considering slowly to moderately rotating stars, we found that some form of macroscopic envelope mixing is required to maintain realistic surface abundances of $^{56}$Fe in order to be compliant with measured abundances of the pulsating B-type stars in our galaxy. 
The need for extra envelope mixing was already deduced from asteroseismic modelling of SPB stars relying on adiabatic pulsation mode computations, in order to match their observed and identified dipole frequencies from 4-year {\it Kepler\/} light curves \citep{Pedersen2021}. We have shown that macroscopic mixing in the radiative envelope of modest rotators caused by IGWs can do the job. We also found it to be possible to excite more modes in B-type stars while maintaining $^{56}$Fe surface abundances consistent with spectroscopic observations \citep[e.g.][]{Gies1992,Morel2006,Niemczura2009, Gebruers2021} by adopting IGW envelope mixing with a density dependence of $\sim\rho^{-1/2}$. While we do not rule out any inaccuracies in the monochromatic opacities of $^{56}$Fe and $^{58}$Ni, we conclude that the  accumulation of heavy elements in the region of the iron-opacity bump caused by radiative levitation can lead to more excited modes than in the case where microscopic atomic diffusion is ignored. Hence our results bring theory and observations of stellar oscillations and surface abundances of slowly to moderately rotating B-type pulsators into better agreement.

\begin{acknowledgements}
The authors thank Aaron Dotter for his help with the numerical implementation of microscopic mixing, Dominic Bowman for his comments on the manuscript prior to submission, and the anonymous referee for the constructive comments provided during the peer-review process.
The research leading to these results has received funding from the 
KU\,Leuven Research Council (grant C16/18/005: PARADISE), from the Radboud University BSc Honours programme offered to RR, and from the Research Foundation Flanders (FWO) by means of a PhD scholarship to MM under project No. 11F7120N. JSGM acknowledges funding the French Agence Nationale de la Recherche (ANR), under grant MASSIF (ANR-21-CE31-0018-02).  The computations of the stellar models and their pulsation modes have been possible thanks to HPC resources from the Flemish Centre for Supercomputing (VSC). RHDT acknowledges support from NASA grant 80NSSC20K0515.
CA also acknowledges the support from the European Research Council (ERC) under the Horizon Europe programme (Synergy Grant agreement N$^\circ$101071505: 4D-STAR).
 While partially funded by the European Union, views and opinions expressed are however those of the authors only and do not necessarily reflect those of the European Union or the European Research Council. Neither the European Union nor the granting authority can be held responsible for them. 
 
\end{acknowledgements}
\bibliographystyle{aa}
\bibliography{Rehm2022.bib}
\begin{appendix}
\section{Chemical profiles} \label{ap:chem_prof}
In this appendix, we show also the profiles of the fractional abundances of several heavy elements for a relatively young main sequence star, and a relatively old main sequence star.
\begin{figure}[h]
    \centering    
    \includegraphics[width=\columnwidth]{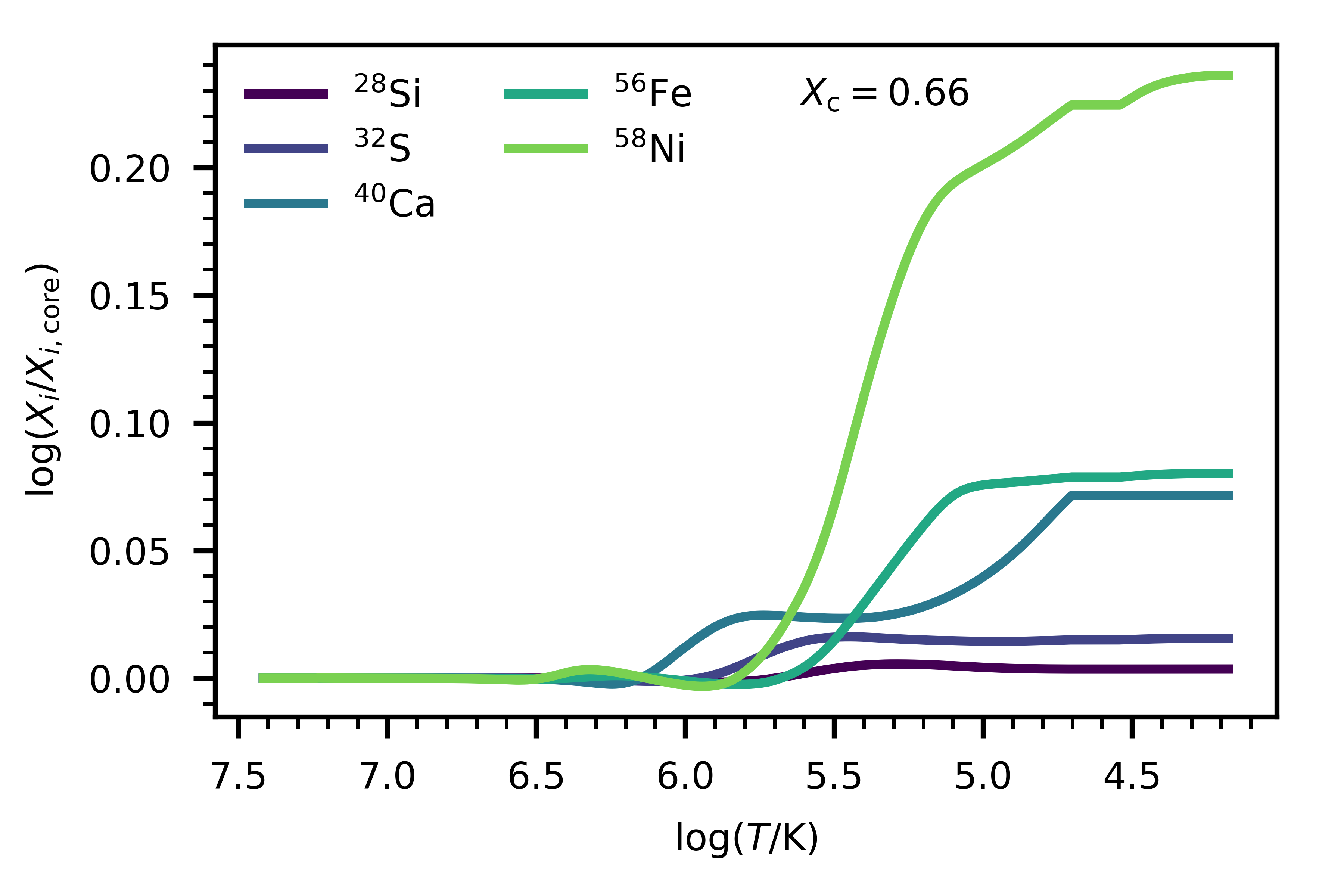}
    \caption{Logarithmic fraction of the local fractional abundance compared to the value in the convective core. This plot shows a model for a 4\Msun star at $X_{\rm c} = 0.66$ ($Z_{\rm{ini}}=0.014$, $\log{D_{0}}=3$, $\alpha_{\rm ov}=0.2$, $\zeta=-0.5$).}
    \label{fig:frac_abun_xc66}
\end{figure}

\begin{figure}[h]
    \centering    
    \includegraphics[width=\columnwidth]{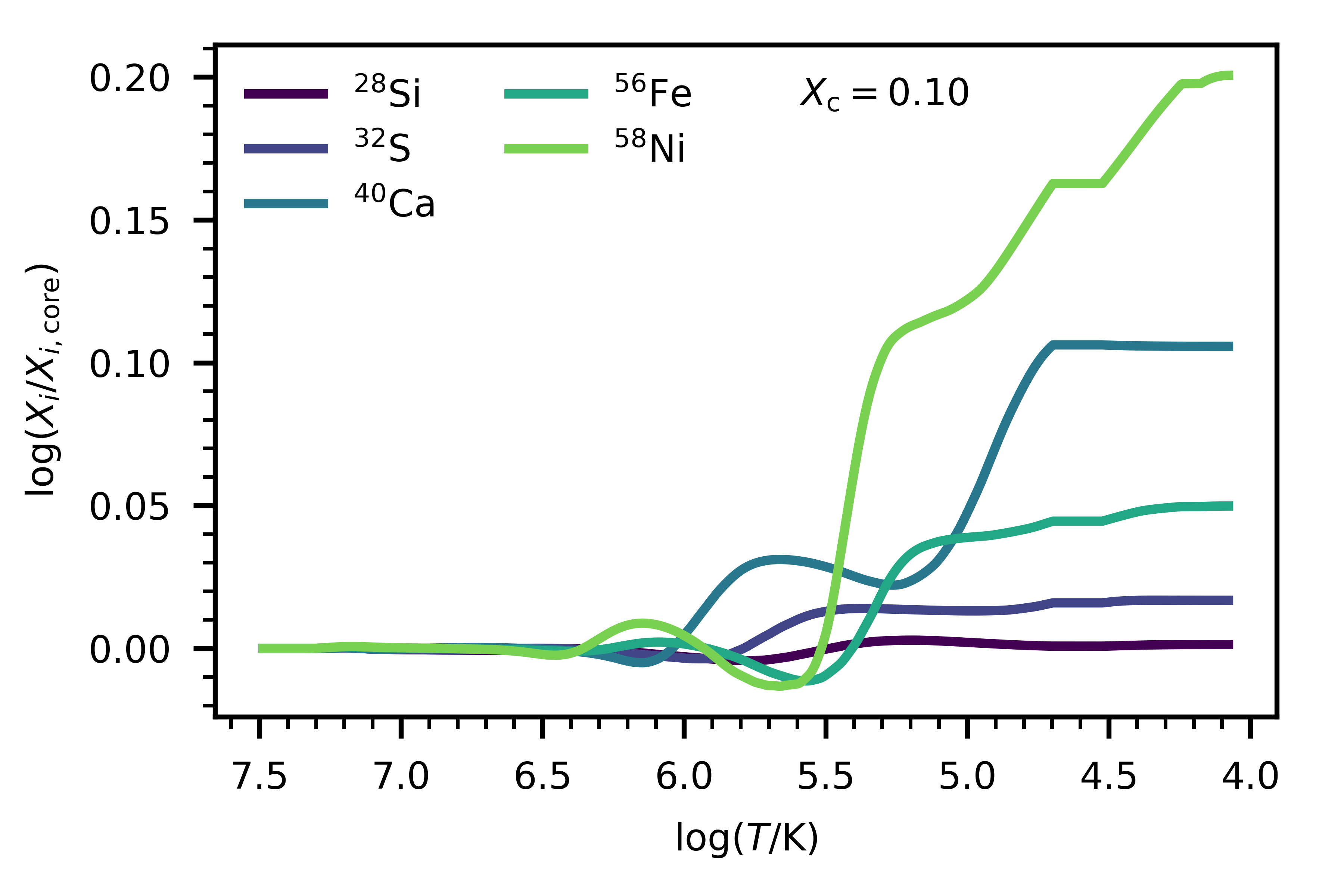}
    \caption{Same as Fig.~\ref{fig:frac_abun_xc66}, but for $X_{\rm c} = 0.10$.}
    \label{fig:enter-label}
\end{figure}

\section{Opacity profiles} \label{ap:opacity}
In this appendix, we also show the local Rosseland mean opacity of the 9\Msun model at different stages throughout the main sequence.
\begin{figure}
    \centering
    \includegraphics[width=\columnwidth]{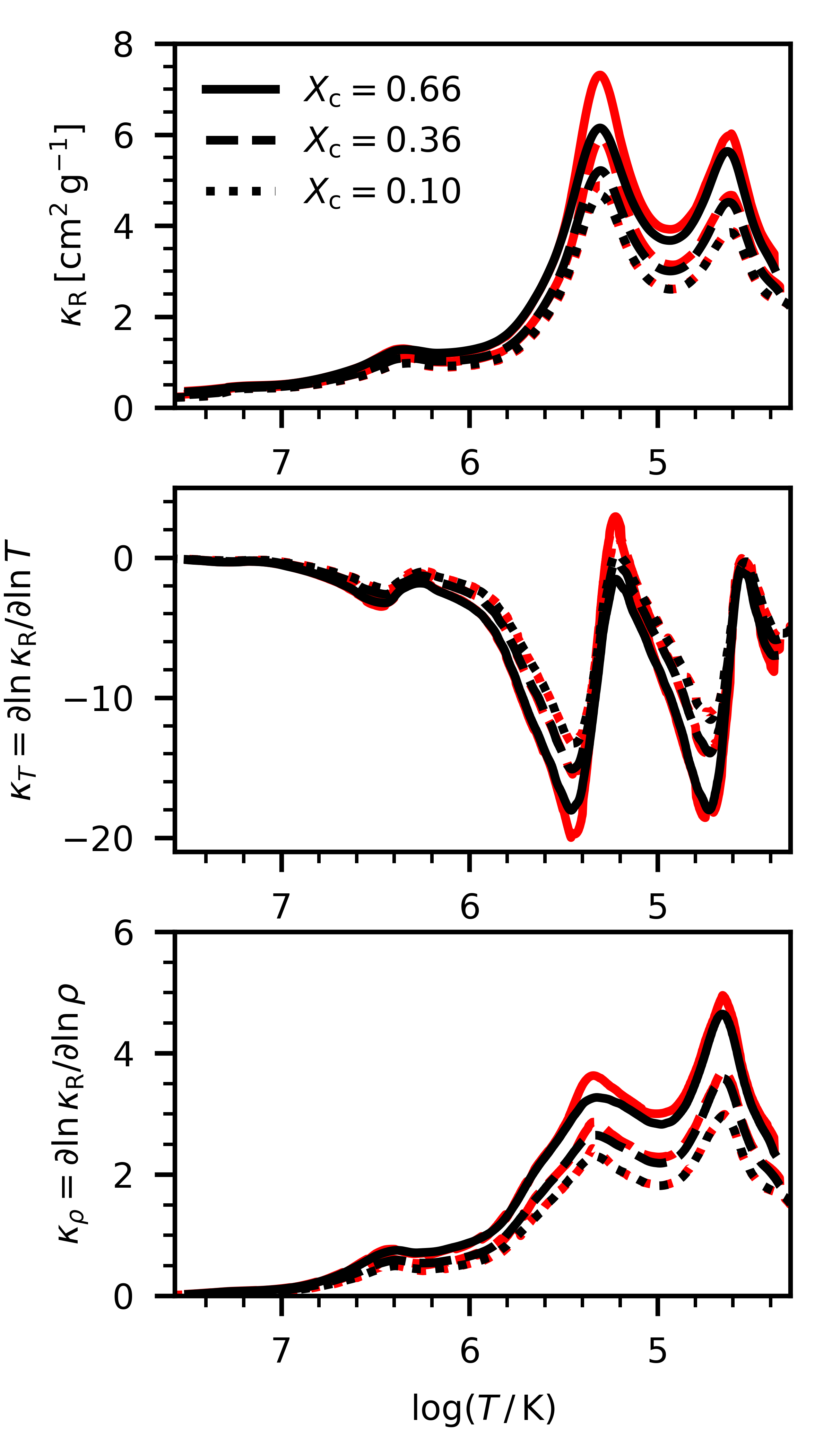}
    \caption{Rosseland mean opacity (top panel), its logarithmic derivative with respect to temperature (middle panel), and its logarithmic derivative with respect to density (bottom panel). Models are shown with diffusion (red) and without diffusion (black) of the 9\Msun model ($Z_{\rm{ini}}=0.014$, $\log{D_{0}}=3$, $\alpha_{\rm ov}=0.2$, $\zeta=-0.5$) at selected $X_{\rm c}$ values.}
    \label{fig:opacity_profile_M9.00_a0.5}
\end{figure}
\end{appendix}

\end{document}